\documentclass{aa}
\usepackage{graphicx}
\usepackage{txfonts}
\usepackage{verbatim}
\usepackage{natbib}
\bibpunct{(}{)}{;}{a}{}{,} 

\begin{document}

\title{Doppler Imaging of LQ~Hya for 1998-2002\thanks{Based on 
observations made with 
the Nordic 
Optical Telescope, operated
on the island of La Palma jointly by Denmark, Finland, Iceland,
Norway, and Sweden, in the Spanish Observatorio del Roque de los
Muchachos of the Instituto de Astrofisica de Canarias. }}

\author{E.M. Cole \inst{1}
\and T. Hackman \inst{1,} \inst{2} 
\and M.~J. K\"apyl\"a \inst{3} 
\and I. Ilyin \inst{4}
\and O. Kochukhov \inst{5}
\and N. Piskunov \inst{5}
}

\institute{Department of Physics, P.O. Box 64 , 
FI-00014 University of Helsinki, Finland \\
\email{Elizabeth.Cole@helsinki.fi}
\and Finnish Centre for Astronomy with ESO, University of Turku, 
V\"{a}is\"{a}l\"{a}ntie 20,
FI-21500
Piikki\"{o}, Finland,
\and ReSoLVE Centre of Excellence, Department of Computer 
Science, PO Box 15400, FI-00076 Aalto University, Finland
\and Leibniz-Institut f\"{u}r Astrophysik Potsdam, An der Sternwarte 16, 14882 Potsdam,
 Germany
\and Department of Physics and Astronomy, Uppsala University, Box 516, 
SE-751 20 Uppsala, Sweden
}
\date{Received / accepted}

\abstract{} 
{We study the spot distribution on the surface of LQ~Hya during 
the observing seasons
October 1998 -- November 2002. We look for persistent active 
longitudes, trends in the level of spot activity
and compare to photometric data, 
specifically to the derived time epochs of the lightcurve minima.}
{We apply the Doppler Imaging technique 
on photospheric spectral lines using an inversion code to 
retrieve images of the surface temperature.}
{We present new temperature maps
using multiple spectral lines for a total of 7 seasons. 
We calculate spot coverage fractions from each map, and as a 
result we find a general trend that is interpreted to be an indication of 
a spot cycle. There is a minimum during the observing season of March 
1999. After this the activity increases until November
2000, followed by a general decrease in activity again.}
{We find no evidence for active longitudes persisting over multiple 
observing seasons. 
The spot activity appears to be concentrated to two latitude regions.
The high latitude spots are particularly strong when the
spot coverage is at a maximum. 
Using the currently accepted rotation period, we find spot structures to
show a trend in the phase-time plot, indicative of a need for a longer period.
We conclude that the long-term activity 
of LQ~Hya is more chaotic than 
that of some magnetically active binary stars analyzed with similar methods, 
but still with clear indications of an activity cycle.
}


\maketitle

\section{Introduction}

\object{LQ~Hya} (\object{HD 82558}, \object{GL 355}) is a young, chromospherically active 
BY~Dra-type star with a spectral classification of K2V 
\citep{Cutispoto91}. 
BY~Dra-type stars are typically
K or M class with strong \ion{Ca}{II} H \& K emission  lines.
LQ~Hya
is considered a young solar analogue, with an estimated 
mass of $0.8 \pm 0.1M_{\odot}$
and an age of $51.9 \pm 17.5$ Myrs \citep{TNH11}. It is a rapid rotator,
with a rotation period $\sim 1.6$ days. It is located 
at a distance of about 18.3 pc.

Flare activity on LQ~Hya has been studied by \cite{CPTP01}, \cite{MSC99}, 
and 
\cite{AF90}. These observations show that
LQ Hya has a high level of 
chromospheric activity. 
LQ~Hya is a single star, as evidenced by a constant radial velocity 
\citep{FMH86}. Photometric variability was proposed
by \cite{Eggen84}, and later confirmed by
\cite{FBAGP86}, where a variation in 
magnitude of 0.1 was measured along with a rotation period
of $P_{\rm rot}=1 \fd 6603$.

LQ~Hya has slight, if any, surface differential rotation. 
The differential rotation rate 
$k$ is defined to be $\Delta \Omega / \Omega_{\rm 0}$
where $\Omega_{\rm 0}$ is the 
angular rotation velocity at the equator and $\Delta \Omega$ is the difference
between the equatorial and polar region rotation.
Estimates from photometry 
based
on observed variations in the photometric period indicate a differential 
rotation coefficient of the
order of $k = 0.025$ \citep{You07},
while estimates from Doppler imaging (hereafter referred to as DI) 
are even smaller, where $k = 0.0057$ \citep{KSGWO04}. 
An even smaller value for $k$ of 0.002 was obtained from Zeeman 
Doppler imaging (hereafter ZDI) 
results \citep{DCP03}.
For comparison,
surface differential rotation in the Sun is $k \approx 0.2$. 
With only small
differential rotation, the dynamo is expected to be of $\alpha^{2}$-type, 
or $\alpha^{2}\Omega$-type where magnetic field generation is dominated
by turbulent convective effects, and any $\Omega$ components from 
the differential rotation are minor \citep{KR80}.

Possible activity cycle periods have been recovered from photometry by 
several groups ranging from 3.2 years \citep{MG03} to up to 12.4 years 
\citep{OKGSL09}. The most commonly reported cyclicity is of 6--7 years
\citep{Jetsu93,SBCR97,Cutispoto98b,OKS00,Alekseev2003,MG03}.

\cite{Jetsu93} found evidence of short-lived active longitudes. When two
such active longitudes were present, they were found to be a phase of 
$\phi=0.25$ apart.  
\cite{BPT02} used 20 years of photometric data, and reported a 
persistent non-axissymmetric
starspot distribution consistent with two alternating
active longitudes with a phase 
difference of 
$\phi=0.5$.
The timescale of this flip-flop phenomenon was claimed
to be of the order of $5.2$ yrs.
Such observations are consistent with observations of other rapid rotators,
such as II~Peg \citep{BT98,LMOPH13} and AB~Dor \citep{JBTCB05}.  
More recently, 
\cite{LJHKH12} found no such stable active longitudes for 
longer timescales 
than 6 months, with the exception for years between 2003-2009 and before
1985. Possible flip-flop phenomena in the light curve were reported
where the main photometric minima switched phase by $\phi=0.5$ with
the weaker secondary minima. 
Similarly, \cite{OKPCH14}
found
no active longitudes in the photometry, 
and coherent structures were short-lived, with the exception of 2005-2008.

There are DI temperature maps 
of \object{LQ Hya}
for January 1991
\citep{SRWHM93}, March 1995 \citep{RS98},
and November-December 1996 and  April-May 2000 \citep{KSGWO04}.
The maps by \cite{SRWHM93} 
were retrieved from single-line inversion solutions, and 
show a spot temperature of $\Delta T_{\rm{spot}} = 500$ K cooler 
than the unspotted surface.
They reported spots at both mid- to low-latitudes and some polar features, but
the appearance of the
latter was highly dependent on the strength of the spectral 
line used for the inversion. 
\cite{RS98} made use of 
multiple spectral lines in the same inversion, and
found bands of 
features around the equator with $\Delta T_{\rm{spot}} = 600$K 
and a weaker polar spot than that previously reported
by \cite{SRWHM93}. The band of spots occupied lower latitudes 
and it was postulated that it coincided with the star being 
at a less active state in the activity cycle than for the January 1991 
maps, although the spot temperature difference was greater. 
The two seasons of maps calculated by \cite{KSGWO04} 
reveal 
changes probably related to the spot cycle.
They retrieved
no polar spot, and spots were largely confined to mid- 
and low-latitudes.

ZDI of LQ~Hya 
ranges from December 1991 \citep{Donati99}
to December 2001 \citep{DCSHP03}.
ZDI maps of LQ~Hya 
over a 5-year period by \cite{Donati99} revealed a shifting spot 
structure and the
concentration of the magnetic field had only a weak 
correlation with respective brightness maps. Another 5-year period was 
covered by \cite{DCSHP03} to construct an 11-year time-series 
of ZDI results. \cite{MJCWD04} selected 
two years of dense observation coverage and found that 
the magnetic field 
topology
differences between high and low activity
states of 
LQ~Hya were
quite different from magnetic field structures at 
solar maximum and minimum. 
Near maximum, in 2000, LQ~Hya 
resembled a tilted dipole, with two open field emergences at 
mid-latitudes separated by approximately $\phi = 0.5$. A year later, 
in December 2001, the field resembled an aligned dipole with 
contributions from east to west arcades. \cite{MJCWD04} concluded that the
magnetic structure underwent rapid changes in less than a year.

\section{Observations}
\label{observations}

\begin{table}
\caption{Summary of observations. 
The name of the image gives the month and year of the start of the 
observing season,
$t_{\rm{min}}$ and $t_{\rm{max}}$ are 
given using the full heliocentric Julian date, S/N is signal-to-noise, 
$n_{\phi}$ is the number of observations, $f_{\phi}$ is the phase coverage, 
and $d$ is the mean deviation of the inversion solution, approximately 
proportional to the inverse of the S/N.}
\centering
\begin{tabular}{lcccccc}
\hline \hline
Image
& $t_{\min}$ & $t_{\max}$ & S/N & $n_\phi$ & $f_\phi$ & $d$\\ 
            & HJD            & HJD         &             &         &    & \\
\hline
Oct98   &  2451089.7    & 2451094.8     & 93      & 6   &  60\%   & 1.01$\%$\\
Mar99   &  2451240.4    & 2451242.6     & 98      & 9   &  57\%   & 1.24$\%$\\
May99  &  2451323.4    & 2451330.4     & 113     & 10  &  72\%   & 0.88$\%$\\
Oct99   &  2451471.8    & 2451475.8     & 117     & 5   &  50\%   & 0.85$\%$\\
Nov00   &  2451854.8   & 2451863.7     & 87       & 7   &  57\%   & 1.20$\%$\\
Feb02   &  2452327.5    & 2452339.6     & 191     & 18   &  80\%   & 0.79$\%$\\
Nov02  &  2452588.8    & 2452606.8     & 249     & 8   &  62\%   & 0.80$\%$\\
\hline
\end{tabular}
\label{obssum}
\end{table}

Our observations were made over a 5-year period using the 
high resolution \'echelle-spectrograph SOFIN at the 2.56 m
Nordic Optical Telescope at
La Palma, Spain. The spectral resolution was $R\approx 70000$. A total 
of 7 observing seasons were used within this time period. 
The observations are summarized in Table \ref{obssum}.
The date, S/N, and phase of each individual observation for each season
are available as online content at the CDS in Table \ref{obsfull}.

Phase coverage $f_\phi$ was quantified by assuming
a phase range of $\phi \pm 0.05$ for each observation. 
\cite{VPH87} studied the robustness of the DI method 
against phase gaps and 
concluded that poor phase coverage may affect spot shape and location, but does 
not introduce spurious spots. This was further tested by \cite{RS00} where large
phase gaps produced warmer starspots and smoother temperature maps. 
\cite{LHMKI14}
examined the changes in observational data by repeating an inversion of a season 
with higher phase coverage, eliminating all but 5, and found the spot 
filling factor increased slightly.
And so we conclude that phase coverage for the temperature maps in
this report is sufficient for drawing some basic conclusions, 
with 50\% or better phase coverage for all seasons
(see Table \ref{obssum}). 
The total observations cover just over 4 years, which is
a few years shy of the oft-cited cycle of about 6-7 years
\citep[e.g.,][]{Jetsu93,Cutipoto98,Alekseev2003}.

The spectral
regions 6172.8 - 6174.3\AA, 6174.7 - 6178.0\AA, and 
6179.6 - 6181.2\AA~ were used for 
the images Oct98, Mar99, May99, and Nov00.
Due to changes in the instrument setup, the regions
6410.9 - 6412.8 \AA, 6419.0 - 6422.5 \AA, and 6430.2 - 6431.8 \AA~ 
were used for Oct99, Feb02, and Nov02. 
To transform the observing times into rotation phases $\phi$, we 
used the ephemeris derived by  \cite{Jetsu93}:

\begin{equation}
\rm{HJD_{min}} = \left(  2445274.22 \pm 0.013 \right) + \left( 1.601136
\pm 0.000013 \right) E.
 \label{ephemeris}
\end{equation}

Later reported rotation periods
are different \citep[e.g.][]{KSGWO04,CMR01,ONSM01}, but
these values are well within the variability in the rotation period determined
by \cite{You07}.

\section{Doppler imaging}
\label{DI}

\begin{table}
\caption{
Adopted parameters
for Individual lines.}
\centering
\begin{tabular}{lccc}
\hline \hline
 Element Ion & $\lambda_\mathrm{centr}$ & $\chi_\mathrm{low}$ & $\log(gf)$ \\  
                   &  (\AA)                     & (eV)              &                  \\
\hline
\ion{S}{1} & 6172.8210 &   8.0450 &   \textbf{-2.400} \\ 
\ion{Fe}{I} & 6173.0080 &   0.9900 &  -7.794 \\
\ion{Eu}{II} & 6173.0290 &   1.3200 &   -0.860 \\ 
\ion{Fe}{I} & 6173.0290 &   3.6400 &   -4.961 \\ 
\ion{Fe}{I} & 6173.3340 &   2.2230 &   \textbf{-2.630} \\ 
\ion{S}{I} & 6173.5840 &   8.0450 &   -1.370 \\ 
\ion{Fe}{I} & 6173.6390 &   4.4460 &   \textbf{-3.390} \\ 
\ion{S}{I} & 6173.7450 &   8.0460 &   -1.400 \\ 
\ion{Ti}{I} & 6174.7510 &   2.6620 &   -1.605 \\ 
\ion{Sm}{II} & 6174.9400 &   1.3480 &   -1.370 \\ 
\ion{S}{I} & 6174.9630 &   8.0460 &   -1.400 \\ 
\ion{Co}{I} & 6175.0200 &   3.6870 &   -1.965 \\ 
\ion{Fe}{II} & 6175.1460 &   6.2230 &   -2.086 \\ 
\ion{Ni}{I} & 6175.3600 &   4.0890 &   \textbf{-0.499} \\ 
\ion{V}{I} & 6175.5650 &   2.8780 &   -1.019 \\ 
\ion{Fe}{I} & 6175.7240 &   4.3010 &   -3.603 \\ 
\ion{S }{I} & 6175.8450 &   8.0460 &   -1.090 \\ 
\ion{Ni}{I} & 6175.9120 &   4.1650 &   -2.831 \\ 
\ion{Fe}{I} & 6175.9130 &   5.0640 &   -3.157 \\ 
\ion{Fe}{I} & 6176.1680 &   4.7960 &   -3.368 \\
\ion{Ni}{I} & 6176.8070 &   4.0880 &   \textbf{-0.220} \\ 
\ion{Si}{I} & 6176.8100 &   5.9640 &   -3.246 \\ 
\ion{Cr}{II} & 6176.9810 &   4.7500 &   -2.887 \\ 
\ion{Ni}{I} & 6177.2360 &   1.8260 &   \textbf{-3.800} \\ 
\ion{Co}{I} & 6177.2710 &   2.0420 &   \textbf{-3.635} \\ 
\ion{Fe}{I} & 6177.4280 &   4.4460 &   \textbf{-3.418} \\ 
\ion{Ni}{I} & 6177.5430 &   4.2360 &   \textbf{-2.041} \\ 
\ion{Fe}{I} & 6179.7890 &   5.2730 &   \textbf{-1.505} \\ 
\ion{Sm}{II} & 6179.8280 &   1.2620 &   -1.380 \\ 
\ion{Ni}{I} & 6179.9900 &   4.0890 &   -2.775 \\ 
\ion{Ce}{II} & 6180.0970 &   1.3190 &   -1.340 \\ 
\ion{Ni}{I} & 6180.1550 &   1.9350 &   \textbf{-3.967} \\ 
\ion{Fe}{I} & 6180.2030 &   2.7270 &   \textbf{-2.506} \\ 
\ion{Ti}{I} & 6180.3030 &   3.4090 &   -0.345 \\ 
\ion{Gd}{II} & 6180.4280 &   1.7270 &   -0.910 \\ 
\ion{Fe}{I} & 6180.5250 &   5.0700 &   -2.308 \\ 
\ion{Co}{I} & 6181.0140 &   3.9710 &   -1.210 \\ 
\ion{Sm}{II} & 6181.0480 &   1.6700 &   -1.060 \\ 
\hline
\ion{Fe}{I} & 6411.1060 &   4.7330 &   -1.920 \\ 
\ion{V}{I} & 6411.2760 &   1.9500 &   -2.059 \\
\ion{Cr}{I} & 6411.5370 &   3.8920 &   -2.478 \\ 
\ion{Fe}{I} & 6411.6480 &   3.6540 &   \textbf{-0.555} \\
\ion{Co}{I} & 6411.8840 &   2.5420 &  \textbf{ -2.528} \\ 
\ion{Fe}{I} & 6412.2020 &   2.4530 &   \textbf{-4.433} \\ 
\ion{Ti}{I} & 6419.0890 &   2.1750 &   -1.656 \\ 
\ion{Fe}{I} & 6419.6440 &   3.9430 &   -2.680 \\
\ion{Fe}{I} & 6419.9490 &   4.7330 &   -0.240 \\
\ion{Fe}{I} & 6420.0640 &   4.5800 &   -2.665 \\ 
\ion{Fe}{I} & 6421.3500 &   2.2790 &   \textbf{-2.088} \\ 
\ion{Ni}{I} & 6421.5050 &   4.1650 &   \textbf{-1.090} \\ 
\ion{Co}{I} & 6421.7030 &   4.1100 &   -1.201 \\ 
\ion{Fe}{I} & 6422.0070 &   4.5930 &   -3.340 \\ 
\ion{Co}{I} & 6430.2900 &   4.0490 &   \textbf{-1.828} \\ 
\ion{V}{I} & 6430.4720 &   1.9550 &   -1.000 \\ 
\ion{Si}{I} & 6430.5590 &   6.1250 &   -1.842 \\ 
\ion{Ca}{I} & 6430.7930 &   3.9100 &   -2.129 \\ 
\ion{Fe}{I} & 6430.8450 &   2.1760 &   \textbf{-2.016} \\ 
\ion{Ca}{I} & 6431.0990 &   3.9100 &   -2.606 \\
\ion{V}{I} & 6431.6230 &   1.9500 &   -1.187 \\ 
\hline
\end{tabular}
\label{spctlines}
\end{table}

DI requires a selection of relatively unblended absorption lines. 
Since the S/N was considerably lower than ideal, it was particularly important
to use multiple lines and weight each phase
in our DI procedure to mitigate the effects of noise 
in each inversion. 
The lines were not added, and so there was
no corresponding improvement in $d$.

Stellar model atmospheres were
taken from the MARCS database
\citep{GEEJN08}. 
The main lines are \ion{Fe}{I} and \ion{Ni}{I} absorption lines. 
The full list of spectral lines 
can be found in Table \ref{spctlines}. 

The continuum was determined in two steps. The spectral orders were first 
normalized by a polynomial continuum fit of the third 
degree. 
as part of the standard spectrum reduction.
This procedure
does not take into account line blending and the possible absence of a 
real continuum within a spectral interval. 
An additional continuum
correction for each wavelength interval was made 
by comparing the seasonal average observed
profile and a synthetic line profile. Near-continuum points were
used for a first or second degree polynomial fit to correct the
normalized flux level.

\subsection{Stellar and spectral parameters}

\begin{table}
\caption{Chosen stellar parameters. References are as follows: 
$(1)$ \cite{Jetsu93}, $(2)$ \cite{RS98}, 
$(3)$ \cite{Donati99}, and $(4)$ \cite{KSGWO04}.}
\centering
\begin{tabular}{rlc}
\hline \hline
Parameter  & Value  & Reference \\ 
\hline
Temperature & $T_\mathrm{eff} = 5000$K  & 3 \\
Gravity & $\log g =4.0$   & 2 \\
Inclination & $i = 65$ \degr & 2  \\
Rotation velocity & $ v \sin i = 26.5 $ km/s  & 3 \\
Rotation period & $P = 1 \fd 6001136$ & 1  \\
Metallicity & $\log [M/H]=0$ & 4  \\
Macroturbulence &  $\zeta_{\rm t}=1.5 $ km/s & 2  \\
Microturbulence & $\xi_{\rm t}=0.5 $ km/s & 2  \\
\hline
\end{tabular}
\label{parameters}
\end{table}

A summary of relevant stellar parameters used in this paper are listed 
in Table \ref{parameters}.
The inversion is sensitive to the 
value for $v \sin i$, and so we determined it by using a model 
with no spots and testing
values $25-28$ km/s and taking the value giving the 
smallest deviation from the mean observed line profile.
The best fit was achieved with $v \sin i = 26.5$ km/s. 
With regards to metallicity, 
we use solar values 
and make individual adjustments to the $\log (gf)$ values of 
specific lines, listed in Table \ref{spctlines}. 
Adjustments are minor, and not particularly of interest as the idea 
is to study the variability in the spectral lines and not determine 
stellar parameters to high accuracy.
These minor changes to values are required to reduce 
systematic errors caused by discrepancies between the 
model and the 
observations. 
We use $i = 65 \degr$, but in reality values  
$\pm 10\degr$ have only a minor impact on the results.

Spectral parameters for the model were obtained from the 
Vienna Atomic Line Database \citep{KPRSW99}. A total of 66 spectral 
lines were used for the images Oct98, Mar99, May99, and Nov00. 
80 spectral lines for the images Oct99, Feb02, and Nov02.

\subsection{Inversion procedure}

We use the inversion method developed by \cite{Piskunov91} 
and further described by \cite{LKHTI11}. This method 
uses Tikhonov 
regularization to stabilize the otherwise ill-posed inversion problem.
In order not to extrapolate, we limit the solution to the temperature range of
atmospheric models used in the calculations, in a similar way as described by
\cite{HJT01}. 
The limits imposed in this study restricted the temperature 
to values between $3400-5500$K.
This range comfortably accommodates the observed $T_\mathrm{eff} = 5000$K 
for LQ~Hya and allows for spot temperatures to be at least 
$1000$K cooler than the mean temperature of the star. 
Previous DI results have shown spots as cool as $4200$K \citep{KSGWO04},
$4400$K \citep{RS98}, and $4700$K \citep{SRWHM93} so the temperature 
range should be more than sufficient to accommodate even the coolest spots.

Line profiles were calculated using 
plane-parallel stellar atmosphere 
models and $\log g = 4.0$ for temperatures between 
$3400-5500$K.
The surface grid resolution 
used for the inversion was $40\times80$ in latitude and longitude, 
respectively. The inversion was run for 30 iterations, at which point
a sufficient convergence was reached. 
Table \ref{obssum} shows the convergence values as $d$, which 
should be approximately the inverse of the S/N. 

\section{Results}
\label{results}

\begin{figure}
\begin{center}
\includegraphics[width=3.4in]{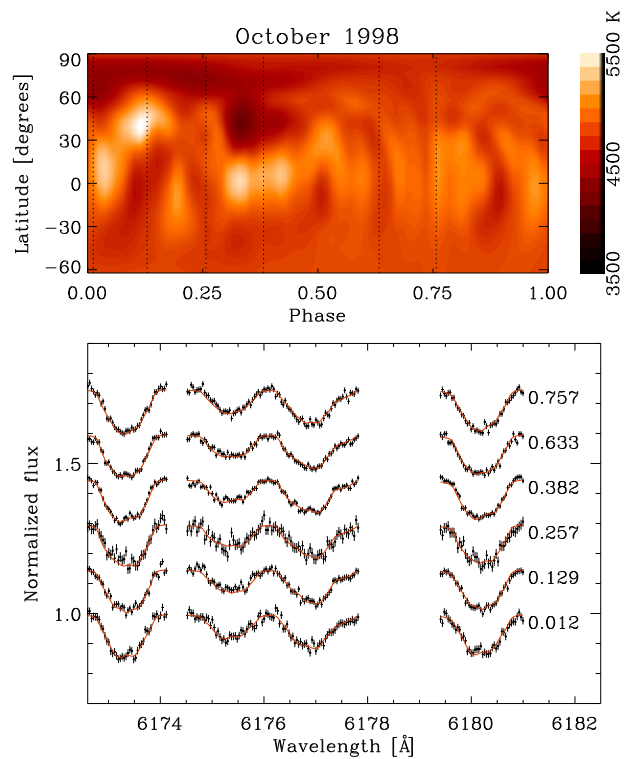}
\caption{{\em Top} Temperature maps in equirectangular
projection. The phase is defined in Eq. \ref{ephemeris}, 
and phases of observations are marked by vertical dashed lines. 
{\em Bottom} Normalized flux of the spectral region used for the inversion.
Points are observations with error, 
the solid red line is the model calculated from the 
DI solution. The phase of each observation is listed to the right 
of the lines.}
\label{Oct98}
\end{center}
\end{figure}

\begin{figure}
\begin{center}
\includegraphics[width=3.4in]{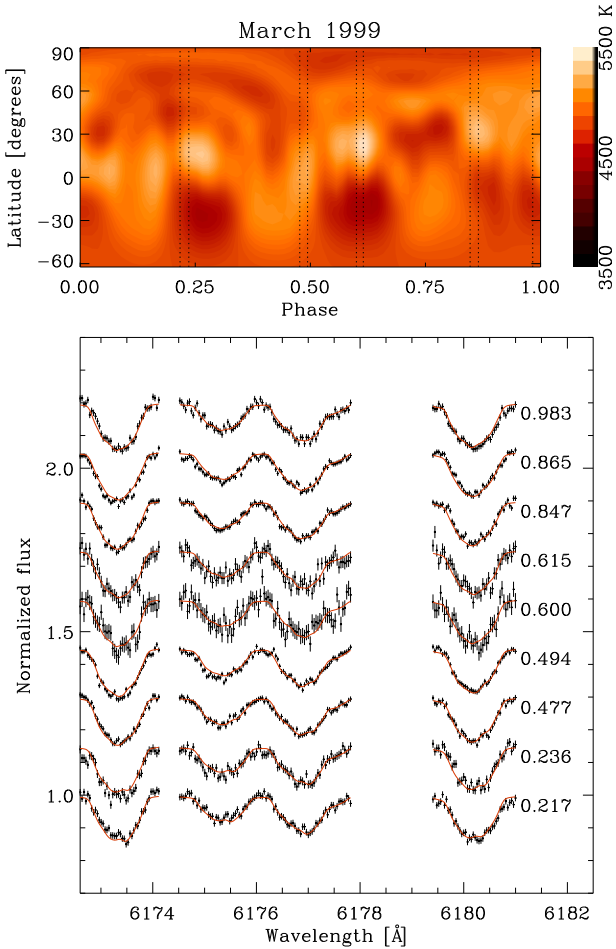}
\caption{Same as in Figure \ref{Oct98}
 for the March 1999 observing season.}
\label{Mar99}
\end{center}
\end{figure}

\begin{figure}
\begin{center}
\includegraphics[width=3.4in]{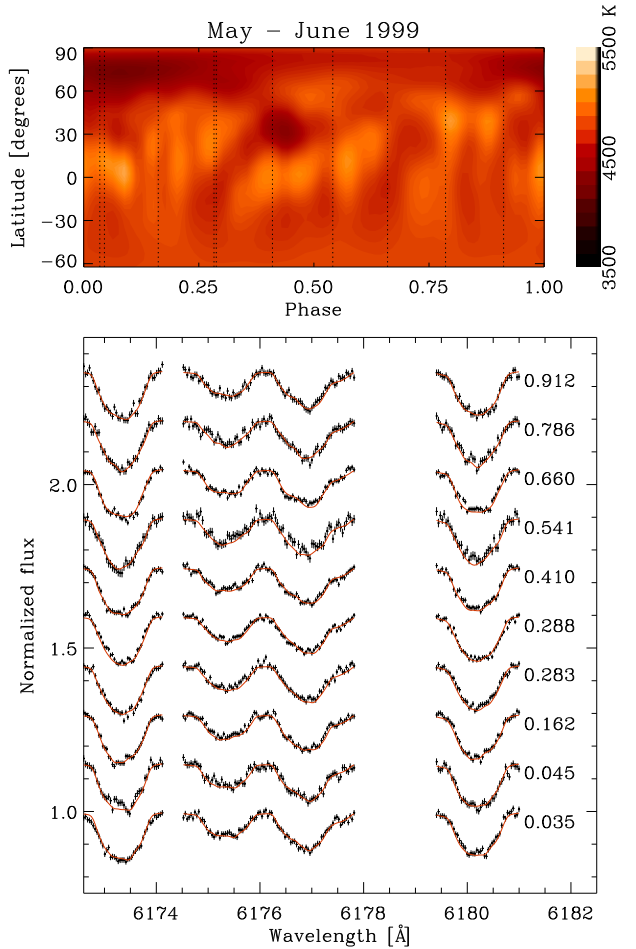}
\caption{Same as in Figure \ref{Oct98} 
for the May - June 1999 observing season.}
\label{May99}
\end{center}
\end{figure}

\begin{figure}
\begin{center}
\includegraphics[width=3.4in]{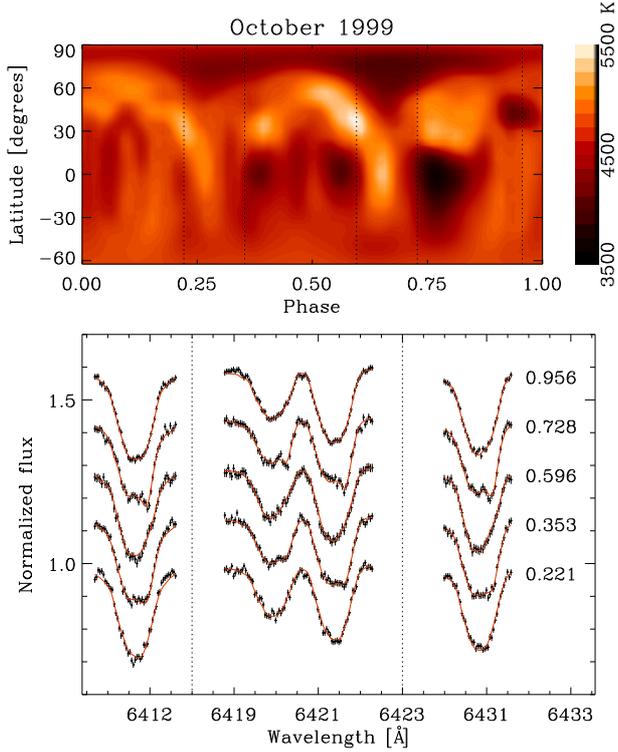}
\caption{Same as in Figure \ref{Oct98}
for the October 1999 observing season. A different 
spectral region was used due to different instrument setup.}
\label{Oct99}
\end{center}
\end{figure}

\begin{figure}
\begin{center}
\includegraphics[width=3.4in]{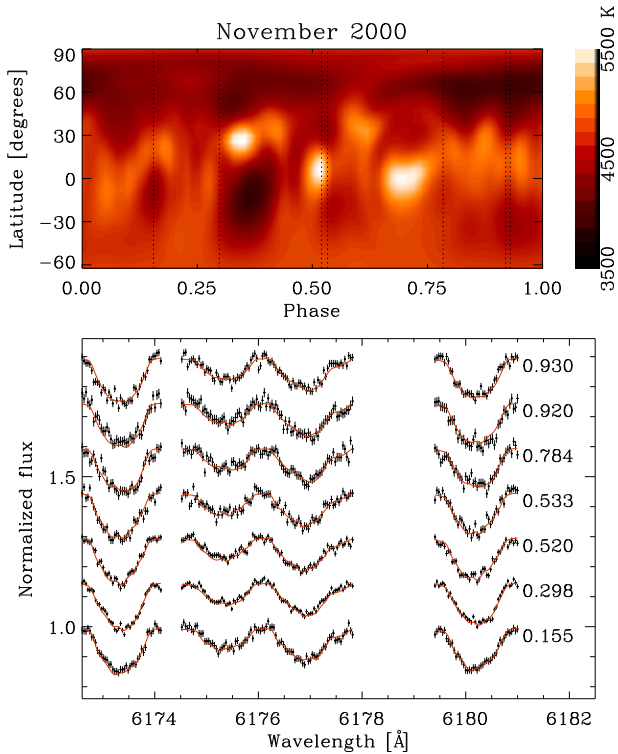}
\caption{Same as in Figure \ref{Oct98} 
for the November 2000 observing season.}
\label{Nov00}
\end{center}
\end{figure}

\begin{figure}
\begin{center}
\includegraphics[width=3.4in]{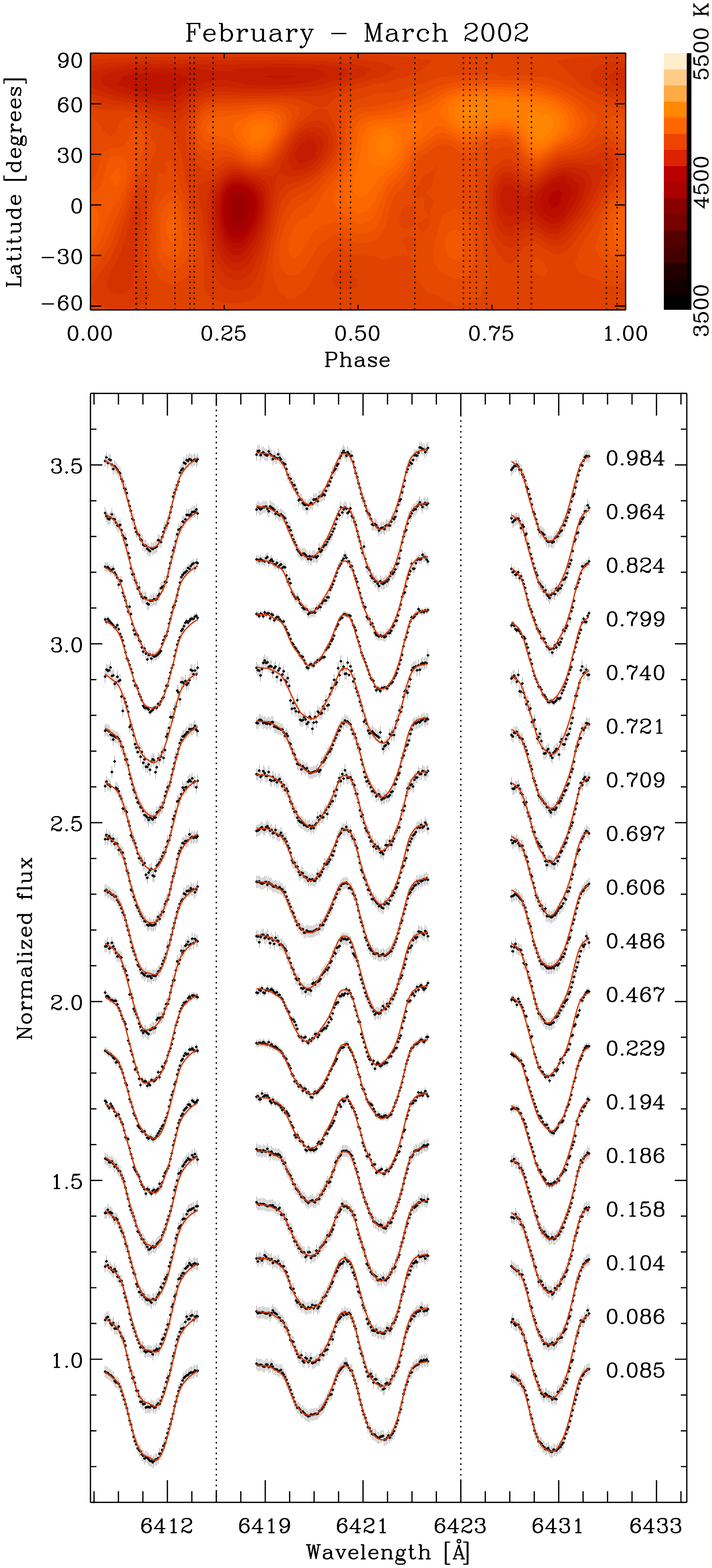}
\caption{Same as in Figure \ref{Oct98} 
for the February-March 2002 observing seasons. 
This is the same spectral region as in Figure \ref{Oct99}}
\label{Feb02}
\end{center}
\end{figure}

\begin{figure}
\begin{center}
\includegraphics[width=3.4in]{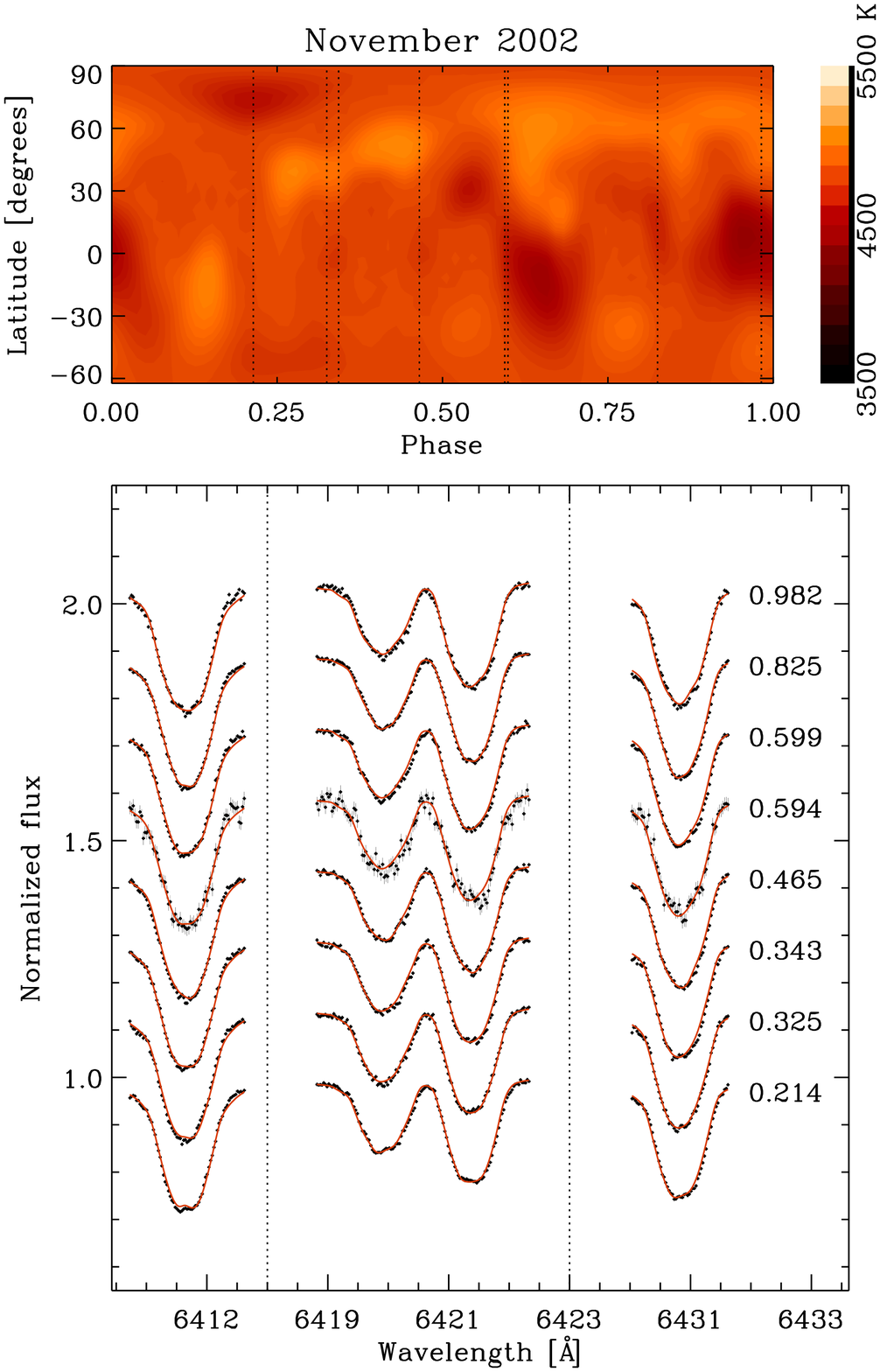}
\caption{Same as in Figure \ref{Oct98}
for the November 2002 observing seasons. 
This is the same spectral region as in Figure \ref{Oct99}}
\label{Nov02}
\end{center}
\end{figure}

The inversions are sensitive to phase coverage and S/N. Lower values 
in either of these for an observing season introduce
artifacts in the temperature maps. 
Certain quantities are more meaningful in these cases than others. 
Phases of spots can be confirmed by examining the 
spectral lines corresponding to each map.
Bumps similarly visible in multiple spectral lines are possible evidence for the
presence of a real spot. 
The mean temperature $\langle T \rangle$ is calculated 
over the visible surface of the 
star and as such, any effects from noise or low phase coverage should cancel 
out with the appearance of both cooler and hotter regions.

However, spot temperature is sensitive to phase coverage and 
noise. Hot regions in the temperature maps 
may be physical, but are also possibly artifacts related either to noise 
or poor phase coverage.
Latitude should also be taken with a grain of salt, particularly with spots
below the equator. The lower the phase coverage, the less information 
available for the inversion program to distinguish latitudes of spots as being 
above or below the equator. 
The inversion method may also interpret sudden changes in spots during 
the observing season as spots at lower latitudes.
A spot at a low latitude would only appear during a limited 
phase range, and sudden spot changes at higher latitudes would have 
similarly limited phase ranges.
For spots close to the equator, noise or errors 
from the continuum level can create artifacts. Arches and ovals that appear 
in the temperature maps are also artifacts, usually related to noise 
or poor phase coverage.
Due to the limitations imposed by the inversion method and the resolution of the 
maps, it is not possible to determine if a spot is a single feature or a composite 
of multiple spots located near each other. 

The Oct98 map (Fig. \ref{Oct98}) shows clear artifacts, 
mainly arches and hot spots in the temperature maps. 
Due to the low phase coverage 
and S/N, the latitude of spots and 
spot temperatures should be viewed with some skepticism. 
Mar99 (Fig. \ref{Mar99}) has slightly 
poorer phase coverage, artifacts, and little evidence for spots within 
the observed phases, as those that appear to
coincide with poorer S/N. May99
(Fig. \ref{May99}) has decent phase coverage and 
S/N, with evenly spaced observations, and minimal appearance of ovals, arches, 
or hot and cool spot pairings. Oct99
(Fig. \ref{Oct99}) has poor phase coverage but S/N is improved. 
Spots at certain phases are distinctly visible in the spectral line profiles as 
consistent deep bumps in multiple lines. 
So while information regarding latitude and spot shape cannot be inferred from this map, 
the phase of spots is likely physical and the amplitude of the bump 
supports a large temperature different between the spot and  
$\langle T \rangle$. 
Nov00 (Fig. \ref{Nov00}) has a lower S/N and poorer phase coverage 
and shows similar evidence of artifacts as Oct99.
However, spot phase and mean temperature are still meaningful due to the proximity of
spots to the observed phases, but unlike the previous season, the noisier spectra 
make it difficult to determine if the spot temperature is due to artifacts or 
physical.
Feb02 (Fig. \ref{Feb02}) has the best phase coverage of 
all the observing seasons, and Nov02 (Fig. \ref{Nov02})  
has the highest S/N.
These two are therefore the most reliable maps.

\section{Discussion}

\begin{figure*}
\begin{center}
\includegraphics[width=3.4in]{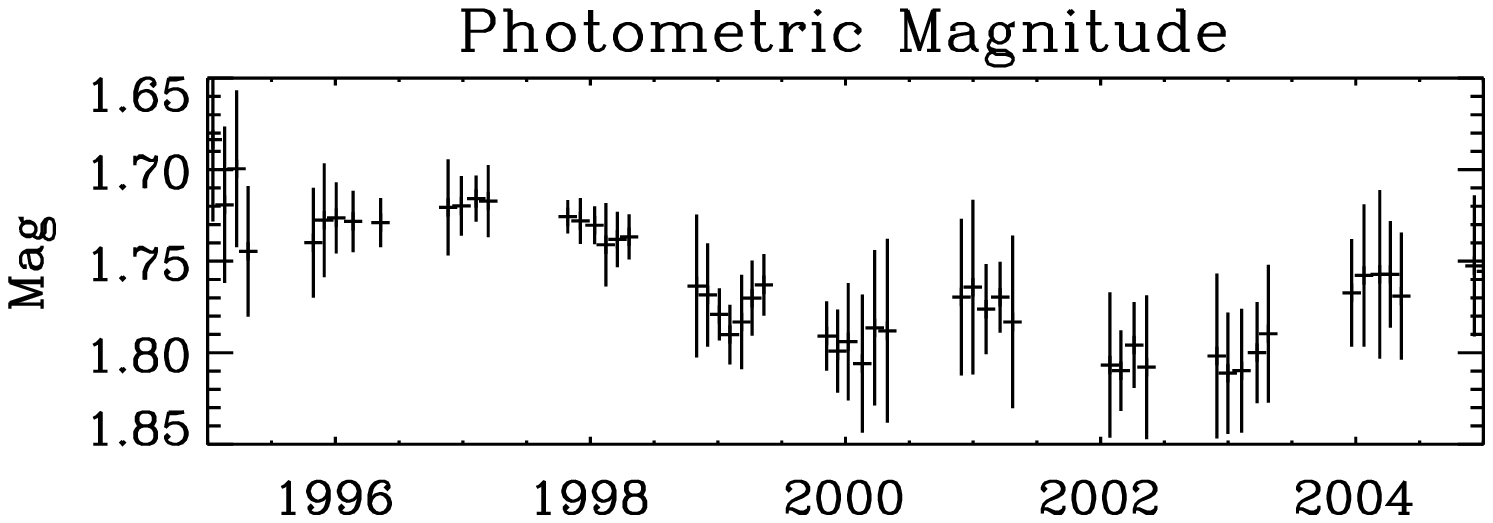}\includegraphics[width=3.4in]{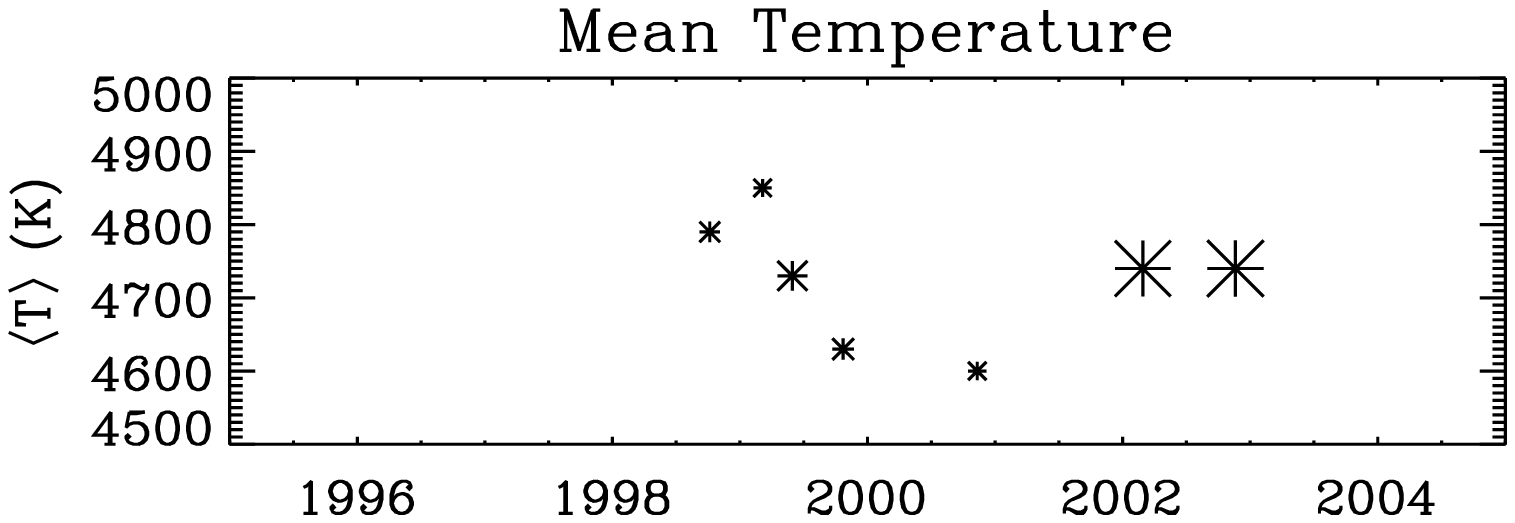}
\includegraphics[width=3.4in]{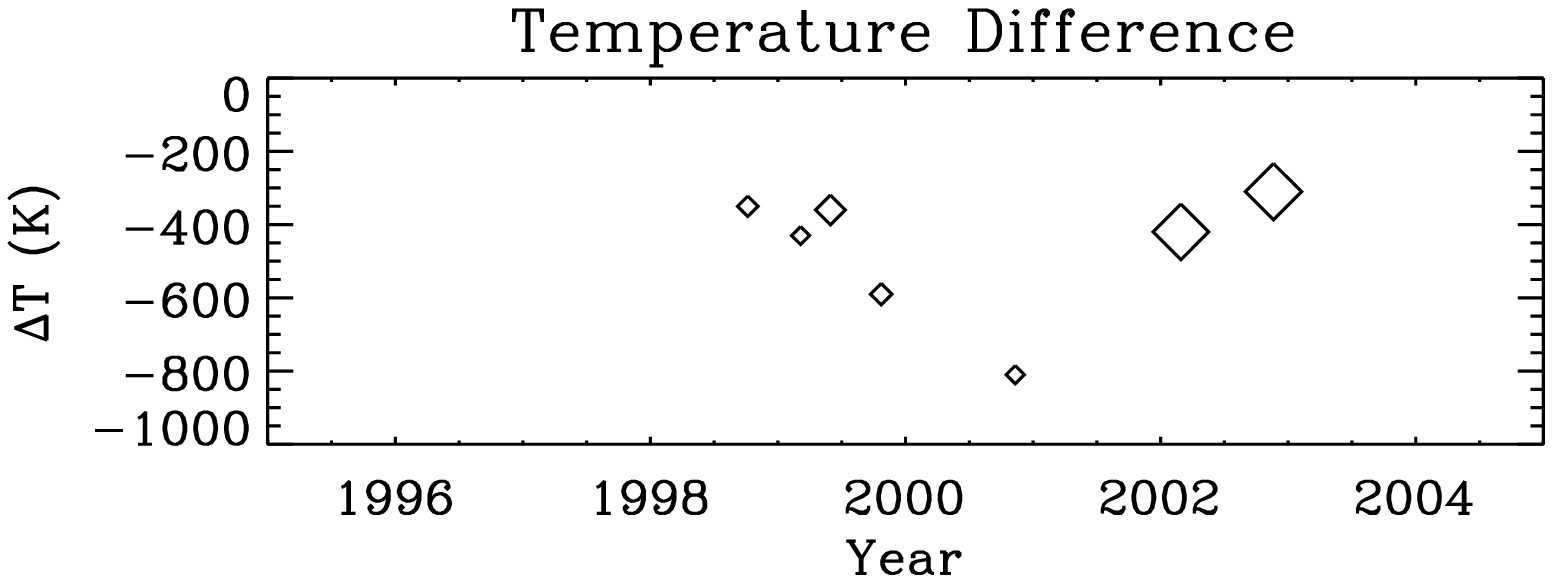}\includegraphics[width=3.4in]{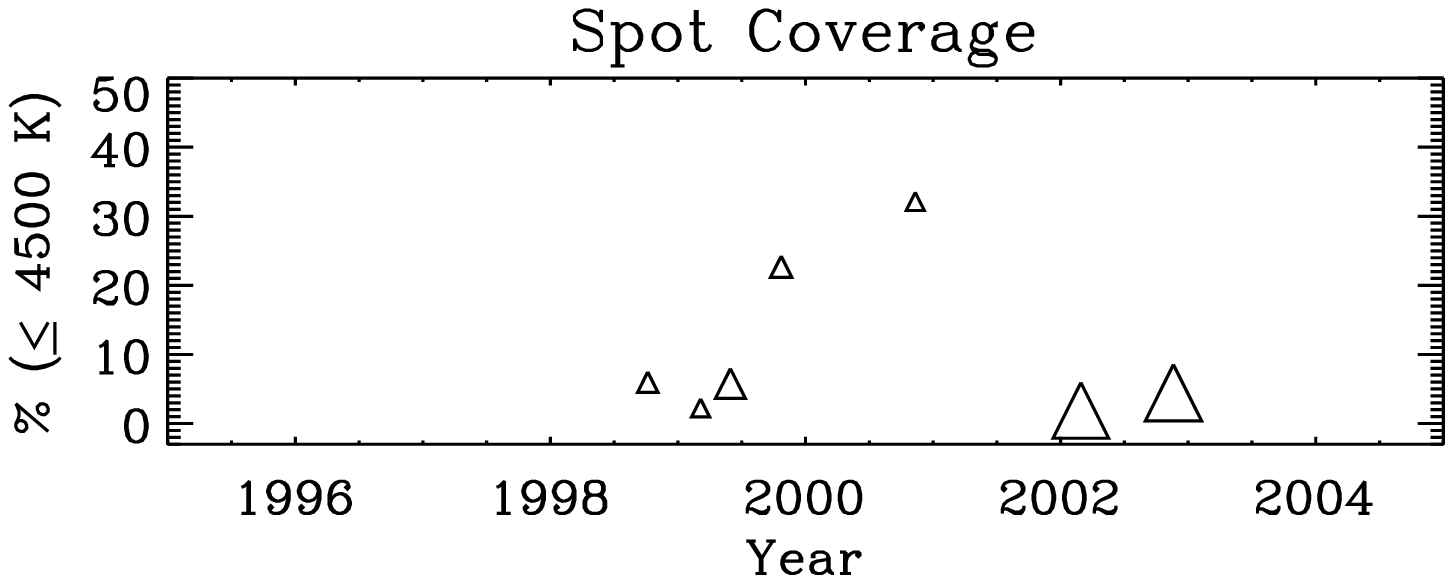}
\caption{Magnitudes are taken from \cite{LJHKH12cat}. Mean temperature, 
temperature difference, and spot coverage. Symbol sizes are 
proportional to the S/N-ratio and phase coverage so that a larger 
symbol indicates more reliable results.
}
\label{summaps}
\end{center}
\end{figure*}

\begin{figure*}
\begin{center}
\includegraphics[width=6.90in]{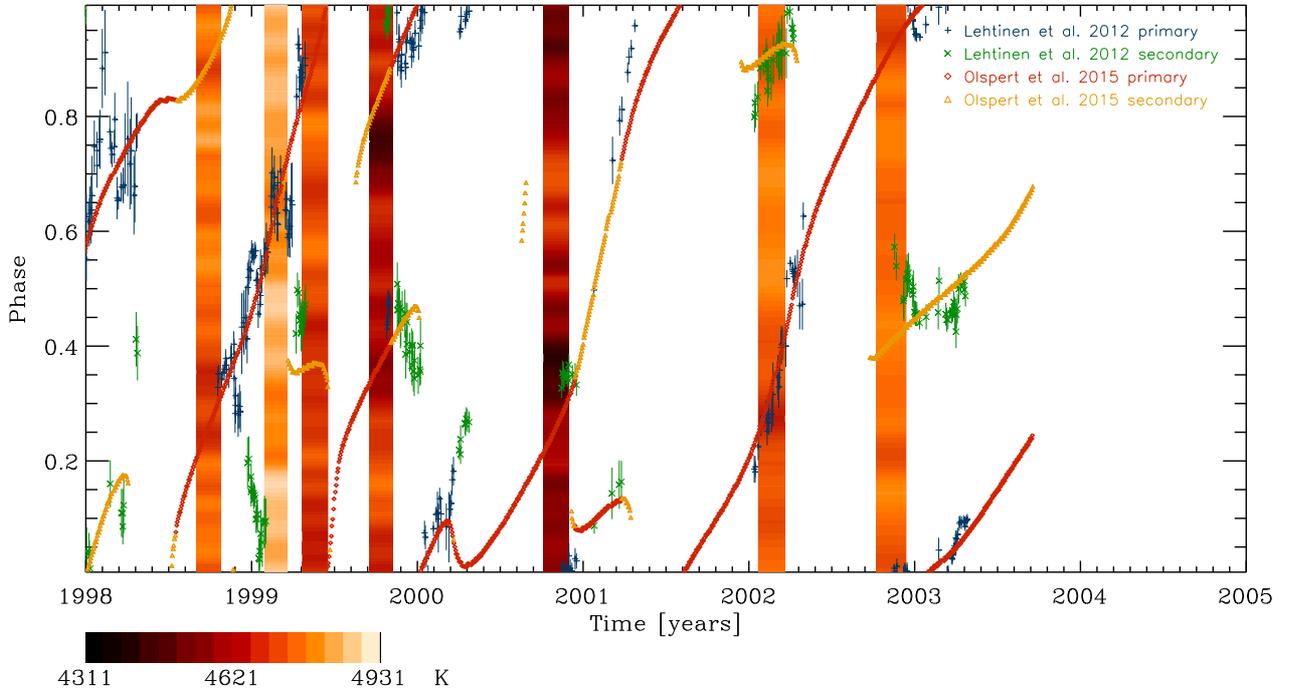}
\caption{Longitudinal spot distribution from October 1998 to November 2002. 
Temperature is averaged over all latitudes for the Doppler Images. 
Also plotted are the photometric minimum epochs 
with error bars from \cite{LJHKH12cat}. 
Minimum epochs were obtained using the CPS method. For a more in-depth 
explanation, see \cite{LJHKH12}.
Minima epochs from \cite{OKPCH14} were obtained using the 
CF method.}
\label{phasetime}
\end{center}
\end{figure*}

\begin{table}
\caption{Spot Coverage, mean temperature, and spot temperature difference 
from the mean.}
\centering
\begin{tabular}{lcccc}
\hline \hline
Season & $n_\phi$ & Spot Coverage & $\langle T \rangle$ & $\Delta T$\\  
\hline
Oct98 &  60\% &  6.0\% & 4790 & 350\\
Mar99 &  57\% & 2.3\% & 4850 & 430\\
May99 &  72\% & 5.8\% & 4730 & 360\\
Oct99 &  50\% & 22.7\% & 4630 & 590\\
Nov00 &  57\% & 32.2\% & 4600 & 810\\
Feb02 &  80\% & 1.9\% & 4740 & 420\\
Nov02 &  62\% & 4.5\% & 4740 & 310\\
\hline
\end{tabular}
\label{spff}
\end{table}

\begin{table}
\caption{Observing season February-March 2002 with original and 
reduced phase coverage obtained by selecting 5 out of 18 
observations.}
\centering
\begin{tabular}{lccc}
\hline \hline
$f_\phi$ & Spot Coverage & $\langle T \rangle$ & $\Delta T$\\ 
\hline
80\% & 1.9\% & 4740 & 420\\
50\% & 9.3\% & 4710 & 700\\
\hline
\end{tabular}
\label{reduce_phase}
\end{table}

For the spot coverage analysis, 
all surface elements having $T_{\rm eff} \le 4500$K were regarded as spots.
Table \ref{spff} contains the mean temperature, coolest spot 
temperature, and spot coverage for each season.
Figure \ref{summaps} illustrates these values in relation to the photometry. 
Results that are less certain due to either low S/N or phase coverage
are represented by smaller symbols. 
We tested the robustness of the inversion 
against phase gaps
by taking the observing season with 
the best phase coverage, February-March 2002,  
and use only 5 observations from the 
18 available. Table \ref{reduce_phase} is the comparison of these two maps. 
While $\langle T \rangle$ is pretty consistent, reducing the phase
coverage increased the spot filling factor from 1.9\% to 9.3\% and
decreased the minimum temperature of the map. Therefore, we can
conclude that the spot filling factor may be overestimated by a
similar factor in the low phase coverage maps.

Spot coverage is low for most seasons, but increases for 
seasons October 1999 and November 2000.
These seasons coincide with an observed decrease in photometric 
magnitude \citep{LJHKH12}.
Even accounting for a low phase coverage, the spot filling factor is
still highest during this point with a corresponding decrease in
$\langle T \rangle$. We could consider this evidence of a possible
cycle, with a high activity state during October 1999 - November 2000.
In the work of \cite{Jetsu93}, a cycle of 6.24 years was obtained from
time series analysis. The ephemeris for the minimum of the mean
brightness was calculated to be
$(1981.^{\rm{y}}49 \pm 0.^{\rm{y}}12) + (6.^{\rm{y}}24 
\pm 0.^{\rm{y}}26)$E. 
This would place the mean minimum brightness, or the highest spot activity, at
$2000.^{\rm{y}}21$, or about March, 2000, which is in agreement with the 
higher spot coverage found during the months preceeding this ephemeris, 
even when accounting for low phase coverage.
 
\begin{figure}
\begin{center}
\includegraphics[width=3.4in]{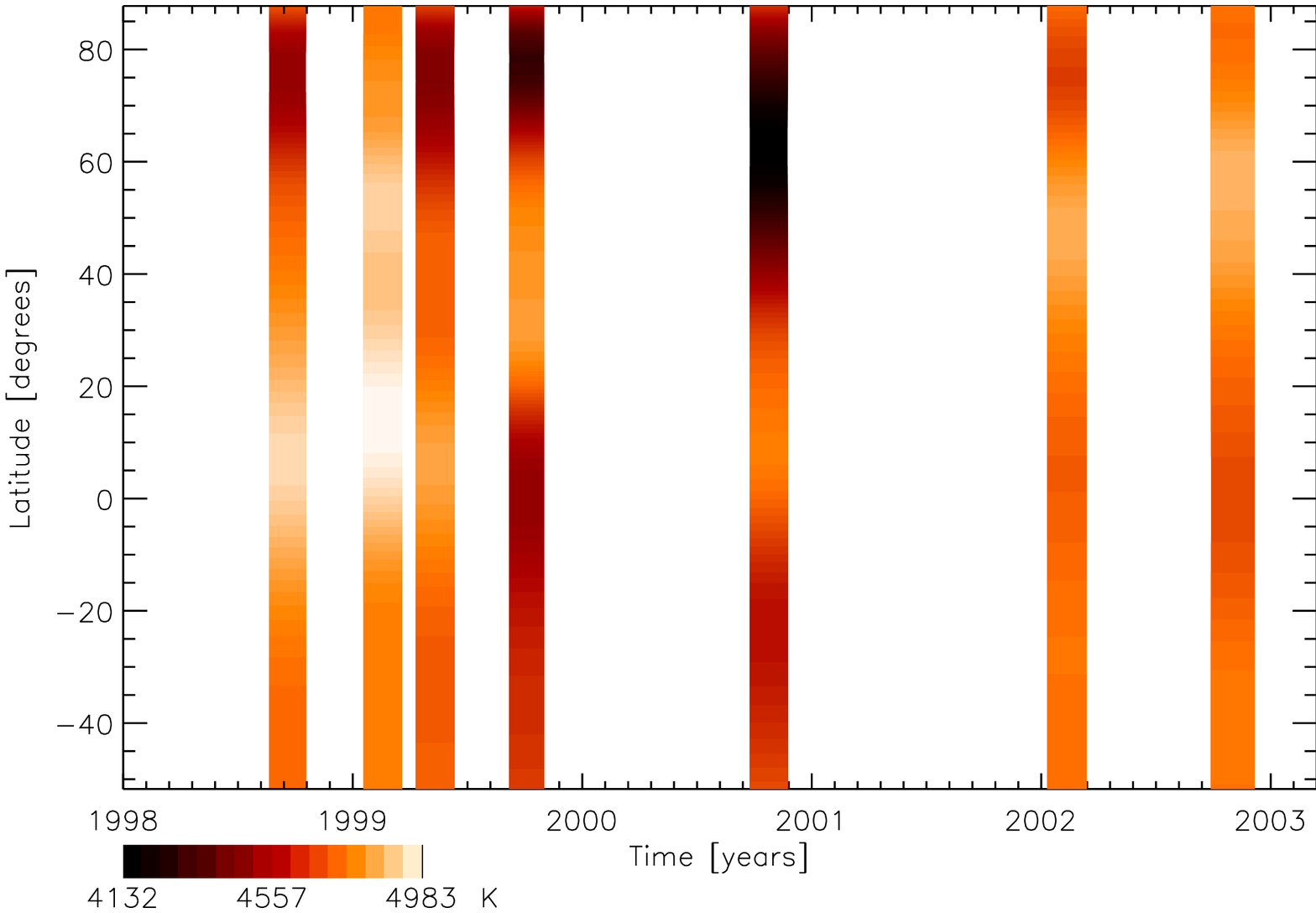}
\caption{Latitudinal spot distribution from October 1998 to November 2002. 
Temperature is averaged over all longitudes for the Doppler Images.}
\label{lattime}
\end{center}
\end{figure}

Figure \ref{phasetime} shows the longitudinal spot distributions by averaging 
over the latitudinal direction. This mimics the results one could 
obtain from photometric observations, where only phase and magnitude 
of the star  at a particular point of time is observed. We have marked the 
primary and secondary minimum epochs retrieved
from light curves from both 
\cite{LJHKH12cat} and from \cite{OKPCH14} using the 
Contiuous Period Search (hereafter CPS) and Carrier Fit 
(hereafter CF) methods respectively. 
Many
seasons show
good agreement with the photometry. For 
instance, in February-March 2002 both the
results from \cite{LJHKH12} and 
\cite{OKPCH14} agree with each other and match
the recovered phases of spot locations in the Doppler images.
The photometric studies coinciding with our DI seasons
reveal an especially chaotic spot activity with non-persistent phase
jumps. 
No coherent active longitudes are seen during this epoch. 
The period used to phase our observations in
longitude is not optimally describing the rotation of the
spot structures, 
evidenced by the upward trend in the global CF fit results.
This is indicative of spot structures moving more 
slowly than the accepted rotation period of the star.
The trend is disrupted at several points, and it would appear that 
the $P_{\rm rot}$ from \cite{Jetsu93} is not coherent for the length of time
of this study.
\cite{OKPCH14} find occurrences of flip-flop type events 
(sudden switches 
in the phase of the primary and secondary minima)
coinciding 
with our observing seasons of October 1999 and November 2000. 
We find no evidence of active longitudes (phases with persistent spots 
over multiple observing seasons), and
our spot phases are not in agreement with the ones 
listed in \cite{BPT02}, and any similar conversion of spot phases
into primary and secondary 
minima reveals no tendency for these spot structures to be 
consistently spaced approximately $180\degr$ apart.

We compare our maps to previous DI results. 
We average the temperature distribution over longitude in Fig. \ref{lattime} 
to facilitate this, keeping in mind that the spot
latitude is less reliable for October 1998, 
October 1999, and November 2000. There is a somewhat bimodal distribution 
with a dark band at or near the 
equator for most seasons as well as a second dark band near the pole. This is 
exaggerated for observing seasons with poorer phase coverage.  

Early DI results by \cite{SRWHM93} for January and February 1991
show spots at mid-latitudes and 
polar features when using specific single spectral lines. 
However, temperature maps from observations during 
March, 1995 by \cite{RS98}
have a band of spots centered around the equator and a slight polar feature. 
The reliable maps of this study show similar weak high-latitude spots, 
but no bands.

Our observations coincide approximately 
with DI by \cite{KSGWO04} and the ZDI by \cite{DCSHP03}, time-wise. 
\cite{KSGWO04} 
observe an increase in spot coverage from 
November and December 1996 to April and May 2000. 
Our temperature maps show a similar increase in the
spot coverage around October 1999.
In contrast to the April and May 2000 maps, our October 1999 and 
November 2000 maps have no band of spots, although there are 
several cool spots located near the equator, but the
latitude is not reliable in these two maps. 

\cite{Donati99} and \cite{DCSHP03} 
published ZDI maps spanning over 10 years.
The spot occupancy results, roughly comparable to 
temperature maps, are sensitive to phase coverage of an observing season. 
Because of this, more reliable seasons have higher spot occupancy and 
so it is difficult to make a comparison between spot occupancy maps and our 
temperature maps with regards to an activity cycle. 
The bimodal distribution of spots near the pole and 
near the equator is in agreement with our results. 

\cite{FMH86} proposed that 
spot evolution for single variable stars would occur more consistently 
and change slower
for variable stars in binary 
systems. 
\cite{HEHH1995} disagreed, postulating that the Roche lobe
would have a stabilizing influence and result in a more consistent 
spot evolution. 
Recent results and this report support the latter conclusion.
Studies of RS~CVn type stars in binary systems such as \object{II Peg} and 
\object{$\sigma$ Gem} seem to display more stable behaviour 
\citep{HMLIK12,KHJLH2014} whereas single type stars such as 
 \object{HD 116956} and FK Com displayed more chaotic behaviour 
\citep{LJHKH2011,HPMJK13}.
LQ ~Hya fits in this latter group, and the DI results 
support this.
There is no appearance of long-lived structures
even within the 4-year time span of maps in this report, 
and while a possible cycle is evident, the spot structure 
evolution over time gives no indication of stability. 

\section{Conclusions}

We have a total of 7 observing seasons from 1998-2002. 
There is a possible cycle with a rise and  
subsequent fall in stellar activity, 
using the spot coverage
and mean temperature
as an indication of 
activity (Fig. \ref{summaps}).
This cycle approximately matches 
a predicted cycle from \cite{Jetsu93}, 
where a decrease in brightness corresponds to an increase 
in spot coverage and therefore activity level.

We compare
the Doppler images to photometry. 
LQ~Hya does not have a consistent 
spot structure over a long period of time, unlike other studied objects 
such as II~Peg. 

There is some evidence for a high-latitude spot, 
but it is not persistent and mainly 
appears during the higher activity seasons. 
There is no evidence of active 
longitudes over multiple observing seasons, 
and this is consistent with 
recent findings using photometry \citep{LJHKH12,OKPCH14}. 
Spots possibly have a bimodal 
distribution (Fig. \ref{lattime}).
and the spot coverage (areas with temperatures $\le 4500$K),
ranges from covering a small amount to a third of the star. 
The cycle length cannot be inferred from only 4 years of observations.
An activity minimum coincides with a 6.24 postulated activity cycle, 
but the baseline of observations falls short of a full cycle length of 6-7 
years. The spot activity is chaotic with flip-flop events in photometry 
occurring during maximum spot coverage.

\begin{acknowledgements}
Financial support from the Academy of Finland Centre of Excellence
ReSoLVE No. 272157 (MJK) and from the Vilho, 
Yrj\"o and Kalle V\"ais\"al\"a Foundation (EC) is gratefully acknowledged.
\end{acknowledgements}

\bibliographystyle{aa}
\bibliography{lqhya1}

\begin{table*}
\caption{All Observations. S/N is calculated is taken from observation
data within the wavelength regions used in 
the analysis. HJD is -2 450 000.}
\centering
\begin{tabular}{cccr|cccr|cccr}
\hline \hline
Date & HJD & $\phi$ & S/N & Date & HJD & $\phi$ & S/N &
 Date & HJD & $\phi$ & S/N \\
(dd/mm/yyyy) & -2450000 & &     &  (dd/mm/yyyy)  & -2450000 & &  & (dd/mm/yyyy) 
& -2450000 & & \\
\hline
03/10/1998	&	1089.7527	&	0.129	&	96	
&	31/05/1999	&	1330.3727	&	0.410	&	127	
&	24/02/2002	&	2329.6036	&	0.486	&	162	\\
04/10/1998	&	1090.7580	&	0.757	&	100	
&	01/06/1999	&	1331.3730	&	0.035	&	116	
&	25/02/2002	&	2330.5620	&	0.085	&	279	\\
05/10/1998	&	1091.7580	&	0.382	&	116	
&	02/06/1999	&	1332.3736	&	0.660	&	128	
&	26/02/2002	&	2331.5424	&	0.697	&	212	\\
06/10/1998	&	1092.7668	&	0.012	&	79	
&	03/06/1999	&	1333.3713	&	0.283	&	120	
&	26/02/2002	&	2331.5807	&	0.721	&	194	\\
07/10/1998	&	1093.7616	&	0.633	&	107	
&	20/10/1999	&	1471.7823	&	0.728	&	91	
&	28/02/2002	&	2333.5716	&	0.964	&	166	\\
08/10/1998	&	1094.7614	&	0.257	&	60	
&	21/10/1999	&	1472.7819	&	0.353	&	127	
&	28/02/2002	&	2333.6028	&	0.984	&	184	\\
02/03/1999	&	1240.3999	&	0.217	&	99	
&	22/10/1999	&	1473.7487	&	0.956	&	179	
&	01/03/2002	&	2334.5984	&	0.606	&	283	\\
02/03/1999	&	1240.4303	&	0.236	&	81	
&	23/10/1999	&	1474.7724	&	0.596	&	121	
&	01/03/2002	&	2335.4832	&	0.158	&	196	\\
03/03/1999	&	1241.4083	&	0.847	&	120	
&	24/10/1999	&	1475.7729	&	0.221	&	107	
&	02/03/2002	&	2335.5272	&	0.186	&	272	\\
03/03/1999	&	1241.4370	&	0.865	&	131	
&	06/11/2000	&	1854.7606	&	0.920	&	63	
&	05/03/2002	&	2338.5694	&	0.086	&	169	\\
04/03/1999	&	1241.6267	&	0.983	&	101	
&	07/11/2000	&	1855.7428	&	0.533	&	80	
&	05/03/2002	&	2338.5981	&	0.104	&	190	\\
04/03/1999	&	1242.4180 &	0.477	&	123	
&	08/11/2000	&	1856.7381	&	0.155	&	85	
&	06/03/2002	&	2339.5679	&	0.709	&	127	\\
04/03/1999	&	1242.4450	&	0.494	&	132	
&	09/11/2000	&	1857.7453	&	0.784	&	69	
&	06/03/2002	&	2339.6174	&	0.740	&	70	\\
05/03/1999	&	1242.6140	&	0.600	&	43	
&	13/11/2000	&	1861.7711	&	0.298	&	125	
&	10/11/2002	&	2588.7586	&	0.343	&	263	\\
05/03/1999	&	1242.6387	&	0.615	&	54	
&	14/11/2000	&	1862.7830	&	0.930	&	83	
&	12/11/2002	&	2590.7615	&	0.594	&	72	\\
24/05/1999	&	1323.3832	&	0.045	&	97	
&	15/11/2000	&	1863.7269	&	0.520	&	104	
&	18/11/2002	&	2596.7350	&	0.325	&	233	\\
26/05/1999	&	1325.3739	&	0.288	&	145	
&	22/02/2002	&	2327.5345	&	0.194	&	165	
&	21/11/2002	&	2599.7598	&	0.214	&	331	\\
27/05/1999	&	1326.3735	&	0.912	&	100	
&	22/02/2002	&	2327.5901	&	0.229	&	245	
&	22/11/2002	&	2600.7389	&	0.825	&	353	\\
28/05/1999	&	1327.3805	&	0.541	&	78	
&	22/02/2002	&	2328.5040	&	0.799	&	201	
&	23/11/2002	&	2601.7633	&	0.465	&	235	\\
29/05/1999	&	1328.3741	&	0.162	&	124	
&	23/02/2002	&	2328.5435	&	0.824	&	162	
&	27/11/2002	&	2605.7937	&	0.982	&	205	\\
30/05/1999	&	1329.3732	&	0.786	&	91	
&	24/02/2002	&	2329.5725	&	0.467	&	156	
&	28/11/2002	&	2606.7807	&	0.599	&	231	\\
\hline
\end{tabular}
\label{obsfull}
\end{table*}

\end{document}